\documentclass[10pt]{article}
\usepackage{graphics}

\begin{document}

\def\oti{{\otimes}}
\def\bra#1{{\langle #1 |  }}
\def\lb{ \left[ }
\def\rb{ \right]  }
\def\tilde{\widetilde}
\def\bar{\overline}
\def\hat{\widehat}
\def\*{\star}

\def\({\left(}		\def\BL{\Bigr(}
\def\){\right)}		\def\BR{\Bigr)}
	\def\BBL{\lb}
	\def\BBR{\rb}
%
%

\def\bb{{\bar{b} }}
\def\ab{{\bar{a} }}
\def\zb{{\bar{z} }}
\def\zbar{{\bar{z} }}
\def\frac#1#2{{#1 \over #2}}
\def\inv#1{{1 \over #1}}
\def\half{{1 \over 2}}
\def\d{\partial}
\def\der#1{{\partial \over \partial #1}}
\def\dd#1#2{{\partial #1 \over \partial #2}}
\def\vev#1{\langle #1 \rangle}
\def\ket#1{ | #1 \rangle}
\def\rvac{\hbox{$\vert 0\rangle$}}
\def\lvac{\hbox{$\langle 0 \vert $}}
\def\2pi{\hbox{$2\pi i$}}
\def\e#1{{\rm e}^{^{\textstyle #1}}}
\def\grad#1{\,\nabla\!_{{#1}}\,}
\def\dsl{\raise.15ex\hbox{/}\kern-.57em\partial}
\def\Dsl{\,\raise.15ex\hbox{/}\mkern-.13.5mu D}
\def\b#1{\mathbf{#1}}
%
%
\def\th{\theta}		\def\Th{\Theta}
\def\ga{\gamma}		\def\Ga{\Gamma}
\def\be{\beta}
\def\al{\alpha}
\def\ep{\epsilon}
\def\vep{\varepsilon}
\def\la{\lambda}	\def\La{\Lambda}
\def\de{\delta}		\def\De{\Delta}
\def\om{\omega}		\def\Om{\Omega}
\def\sig{\sigma}	\def\Sig{\Sigma}
\def\vphi{\varphi}
%
%
\def\CA{{\cal A}}	\def\CB{{\cal B}}	\def\CC{{\cal C}}
\def\CD{{\cal D}}	\def\CE{{\cal E}}	\def\CF{{\cal F}}
\def\CG{{\cal G}}	\def\CH{{\cal H}}	\def\CI{{\cal J}}
\def\CJ{{\cal J}}	\def\CK{{\cal K}}	\def\CL{{\cal L}}

\def\CM{{\cal M}}	\def\CN{{\cal N}}	\def\CO{{\cal O}}
\def\CP{{\cal P}}	\def\CQ{{\cal Q}}	\def\CR{{\cal R}}
\def\CS{{\cal S}}	\def\CT{{\cal T}}	\def\CU{{\cal U}}
\def\CV{{\cal V}}	\def\CW{{\cal W}}	\def\CX{{\cal X}}
\def\CY{{\cal Y}}	\def\CZ{{\cal Z}}

\def\rvac{\hbox{$\vert 0\rangle$}}
\def\lvac{\hbox{$\langle 0 \vert $}}
\def\comm#1#2{ \BBL\ #1\ ,\ #2 \BBR }
\def\2pi{\hbox{$2\pi i$}}
\def\e#1{{\rm e}^{^{\textstyle #1}}}
\def\grad#1{\,\nabla\!_{{#1}}\,}
\def\dsl{\raise.15ex\hbox{/}\kern-.57em\partial}
\def\Dsl{\,\raise.15ex\hbox{/}\mkern-.13.5mu D}
\def\beq{\begin {equation}}
\def\eeq{\end {equation}}

\title{Low-Entanglement Remote State Preparation}
\author { Igor Devetak\footnote{Electronic address: igor@ece.cornell.edu}\, and Toby Berger 
                                        \\
\emph{Department of Electrical and Computer Engineering}
  \\
 \emph{Cornell University, Ithaca, New York 14850} 
}
  \date{\today} 
  \maketitle

\begin{abstract}
  
An outer bound on the low-entanglement remote state preparation (RSP)
ebits vs. bits tradeoff curve \cite{bennett} is found 
using techniques of classical information theory.
We show this bound to be optimal among an
important class of protocols and conjecture optimality even without 
this restriction.

\end{abstract}  
\vspace{2mm}

We all know what state preparation is: Alice, having complete classical knowledge of a quantum state,
prepares it in her lab. 
Remote state preparation (RSP) refers to the case where Alice, again having a classical description of the 
state, wishes to prepare a physical instance of it in Bob's lab, Bob being far away. It seems natural
to ask about how this situation differs from quantum teleporation \cite{tele} where Alice has no classical 
knowledge of state, but has a physical instance of it.
This was first addressed by Pati \cite{pati} and 
Lo \cite{lo} who considered special ensembles of states. The general case was investigated by Bennett et al.
\cite{bennett} where they posed the question of quantifying the resources necessary and sufficient for 
asymptotically perfect RSP. Asymptotic perfection means that the average fidelity between the resulting 
states in Bob's lab and the corresponding states Alice intended him to prepare tends to $1$ as the number of
states to be remotely prepared is taken to infinity. The resources are the same as for teleportation:   
entanglement (ebits) between Alice and Bob and classical bits of forward communication from Alice to Bob. 
They also allow classical back-communication from Bob to Alice, this extra resource being unhelpful
for teleportation. For the case of qubit states Bennett et al. found outer bounds on the achievable (b,e) 
pairs by explicit construction of RSP protocols (see Fig.1). The teleportation point $(2,1)$ naturally divides the 
plane into a high and low-entanglement region where the number of ebits per remotely prepared state is greater
than and less than $1$, respectively; there is a large qualitative difference in the methods used
for these two cases. The high-entanglement region is accessed by Alice performing  certain
generalized measurements on her ebit halves that possibly depend on her classical knowledge of the state, 
and sending classical information about the measurement
results to Bob. The low-entanglement protocols described in \cite{bennett} (which we
refer to as \emph{teleportation based}) involve sending classical 
information about the states themselves causing
a reduction in the posterior von Neumann entropy from Bob's point of view, and teleportation of Schumacher 
compressed states. Here we concentrate on the latter case, pushing these ideas to their information 
theoretical limit. The main result is an analytic expresssion for the best teleportation based outer bound 
on the low-entanglement region.
Our approach borrows heavily from Shannon's classical rate-distortion theory \cite{berger} \cite{cover}, 
and we will emphasize the key concepts and ideas, relegating technical details to a future publication \cite{long}.

\vspace{7mm} 
\centerline{ {\scalebox{0.7}{\includegraphics{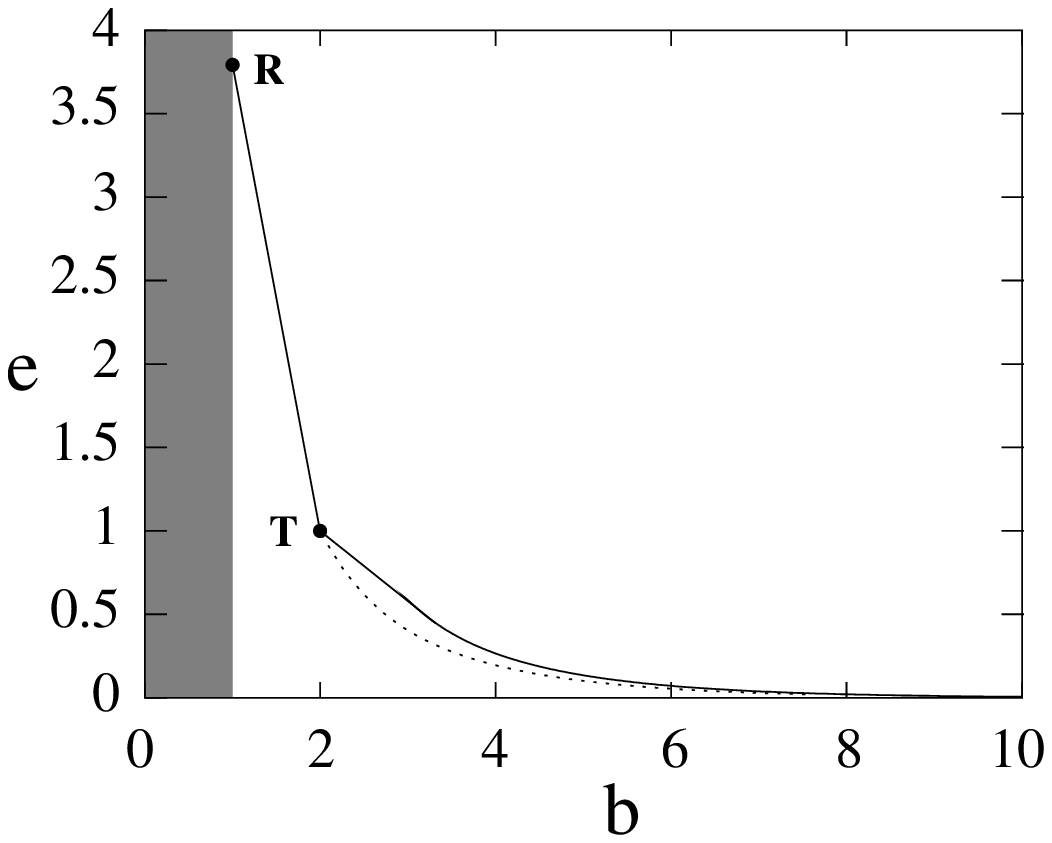}}}}

{\small {FIG. 1} Ebits vs. bits for remote state preparation (from \cite{bennett}).
The dotted curve represents
our low-entanglement outer bound. The solid curve is the previous outer bound
by Bennett et al.
The shaded region is forbidden by causality.}
 
 \vspace{5mm}
 
Let us first consider an example (attributed to H.-K. Lo \cite{bennett} \cite{lo}) illustrating
the way classical information about a qubit state reduces its von Neumann entropy. 
It is important to appreciate the fact that, in the scenario we are dealing with here, the density
matrix is not a property of the qubit, but rather reflects \emph{knowledge} about the actual 
pure state the qubit is in. Alice knows her states \emph{exactly} prior to remotely preparing them,
hence the individual density matrices have zero entropy, from her point of view. At the same time
Bob is completely ignorant of the qubit states; for all he knows Alice could have chosen
them from anywhere on the Bloch sphere.  
More formally, if we denote the Bloch sphere by $\CX$, parametrized by  spherical polar 
coordinates $x \equiv (\theta_x,\phi_x) \in [0, \pi] \times [0, 2 \pi]$
(for convenience we will refer to the north pole
$\theta_x = 0$ as $x = 0$), then 
the probability density corresponding to picking $x$ is simply $p(x) = {1 \over {4 \pi}}$.
The corresponding quantum state is $\ket{x} = \sqrt{\frac{1 + 
\cos\theta_x}{2}} \ket{0} + e^{i \phi_x}
\sqrt{\frac{1 - \cos\theta_x}{2}} \ket{1}$.
The resulting density matrix from Bob's point of view is 
$\rho =  \int d{x} p({x}) \ket{{x}} \bra{{x}} = \frac{1}{2} I $,
and the von Neumann entropy is $S(\rho) = 1$, as one would expect from such a 
random distribution. 
Now, let us assume Alice gives Bob $1$ bit of classical information about
the state, e.g., tells him whether the state is in the upper or lower Bloch hemisphere. 
The \emph{posterior} distribution is now uniform in the upper
(lower) hemisphere,
i.e. $p'(x) = {1 \over {2 \pi}}$ for $x$ in the upper(lower) hemisphere and zero otherwise.
The density matrix $\rho'$ is computed as above, and the posterior von
Neumann entropy becomes $S(\rho') \approx 0.81$ in either case. 
Schumacher's theorem \cite{schum} now tells us that
we have reduced the amount of quantum information to be conveyed to Bob, at the expense of
an additional classical rate of $1$ bit per letter. Based on this observation a protocol may be devised
as follows \cite{bennett}.

\vspace{1mm}

$\bullet$ Alice sends classical information to Bob at a rate $R = 1$ bit per remotely prepared state 
about which hemisphere the state lies in.
 
$\bullet$ This reduces the von Neumann entropy of the source (as viewed by Bob) to
$S \approx 0.81$. However, the density matrices now depend on the hemisphere.
So Alice rotates, say, all the states in the lower hemisphere by a preagreed
unitary transformation that maps the lower onto the upper hemisphere (any rotation
sending the south pole to the north accomplishes this).
Now the qubits are  i.i.d. from Bob's point of view, and Schumacher's theorem applies. Alice prepares these 
rotated states, and Schumacher compresses them to $S$ qubits per letter.
  
$\bullet$ Alice teleports the compressed qubit states at a rate of $2S$ bits and $S$ ebits
per remotely prepared state. 

$\bullet$ Bob simply reverses Alice's steps in his laboratory, thus recovering 
asymptotically faithful instances
of her states.

\vspace{1mm}

This teleportation based protocol yields the point $(2S + R, S)$ in the (b,e)-plane. 
The property of being asymptotically faithful
is inherited from Schumacher compression, this being based on classical Shannon compression.
It is a low-entanglement protocol since $e = S \leq 1$.
It is now evident that, if we restrict attention to teleportation based protocols,
the problem reduces to finding the optimum \emph{rate-entropy curve}, i.e. the
frontier of $(R,S)$ pairs attainable in this way. One may wonder, for example, how it is possible
to further reduce $S$ while keeping $R = 1$. The answer lies in exploiting the asymptotic
formulation of the problem and processing blocks of states, now minimizing 
the entropy per remotely prepared state.

We proceed to formulate the source coding problem. 
The \emph{source} is described by a random vector 
$\mathbf{X} = (X_1,X_2, \dots,X_n)$, and we take the individual $X_i$ to be independent
and identically distributed (i.i.d.), each taking values $x$ on the Bloch sphere $\CX$ 
with probability density $p(x) = {1 \over {4 \pi}}$. 
Thus the probability density distribution
for $\b{X}$ is $p(\b{x}) = \prod_i p(x_i)$.
This reflects Bob's view before he receives
any classical information.
Elements $\b{x} = (x_1,x_2, \dots, x_n)$ of $\CX^n$ are called 
\emph{source words} of length $n$, and the $x_i$ are called \emph{letters}.
We map the source $\mathbf{X}$ onto
a set $B_n = \{ \mathbf{y}_1,\b{y}_2, \dots , \mathbf{y}_K \}$, $\b{y}_k \in \CX^n$,
called a \emph{source code}
of \emph{size} $K$ and \emph{blocklength} $n$, of reproducing \emph{codewords}.
The \emph{rate} of the code is formally defined as $R = n^{-1} \log_2 K$, and it signifies 
the number of bits per source letter needed to specify the index of the reproducing codeword.
When Bob recieves these $R$ bits, he knows the reproducing codeword, 
which  is an approximation to the actual source word. 
In Lo's simple example $n = 1$, $K = 2$, $R = 1$ and  $B_1$ consists of two codewords corresponding
to the north pole $y_1$ and south pole $y_2$, respectively.
There each source word gets mapped onto the closest pole, and knowledge of the codeword
is equivalent to specifying the hemisphere.
The goal is to minimize the von Neumann entropy of the source word 
as viewed by Bob upon receiving the reproducing codeword.
Formally, each source word $\b{x}$ gets mapped into a unique $\b{y} \in B_n$ in such a way that 
the posterior von Neumann entropy of the source
 
\beq
   {S}(B_n) = \frac{1}{n} E_\mathbf{Y} S(E_\mathbf{X|Y} \ket{\mathbf{X}}
   \bra{\mathbf{X}})
  \label{eqn11}
\eeq  

is minimized. Here $\mathbf{Y}$ is the random vector associated with the
probability distribution on the set of codewords $B_n$ induced by our map. 
$E_\mathbf{Y}$ denotes the expectation value over the random vector
$\mathbf{Y}$, and $E_\mathbf{X|Y}$ is the conditional expectation over 
$\mathbf{X}$  given the value of $\mathbf{Y}$. Let us analyze the above expression.
Let $\CM_{\b{y}}$ be the set of all values of $\b{X}$ that get mapped into $\b{Y} = \b{y}$.
When Bob learns that  $\b{Y} = \b{y}$ he knows that $\b{X}$ must have come from
the set $\CM_{\b{y}}$. The density matrix he sees is now an average over all the 
$\b{X}$'s from $\CM_{\b{y}}$ and is denoted by the expectation value $E_\mathbf{X|Y=y}
\ket{\mathbf{X}}\bra{\mathbf{X}}$. We average
the corresponding von Neumann entropy over all the possible $\mathbf{Y}$'s Bob could have received, 
and divide by $n$ to get a per letter result, thus giving rise to (\ref{eqn11}).
In Lo's  example the random variable $Y$  takes
on the values $y_1$ and $y_2$ with probabilities $\frac{1}{2}$ each, depending on the 
hemisphere of $X$. The distribution of $X$ given $Y$ is uniform over the 
hemisphere indicated by the value of $Y$. Thus (\ref{eqn11}) indeed  
yields the entropy obtained before.

Formally, a rate-entropy pair $(R,S)$ is called (asymptotically) achievable iff there exists a sequence of
source codes $B_n$ of rate $R$ and increasing blocklength $n$ such that

       \beq
      \lim_{n \rightarrow \infty} S(B_n) \leq S
       \eeq 
   
We now define the rate-entropy function $R(S)$ as the infimum of
all $R$ for which $(R,S)$ is achievable. 
The way such a coding problem can be solved exactly is by first
finding an information-theoretical lower bound on $R(S)$ and then producing a coding scheme 
that achieves said bound. Firstly, note that $\b{Y}$ is completely determined by the corresponding 
value of $\b{X}$, and hence the conditional probability density
$Q(\b{y}|\b{x})$ is a $\delta$-function. However, for the purpose of finding a lower
bound we relax this constraint. 
Secondly, observe the following string of inequalities

\beq
     R = {\frac{1}{n}} \log_2 K  
     \geq {\frac{1}{n} H(\b{Y})}
     \geq \frac{1}{n} I(\b{X};\b{Y}) 
     \label{eqn2}
\eeq
 
The first inequality is saying that the entropy of $\b{Y}$ is maximum when the codewords
occur with equal probability $K^{-1}$ in which case the entropy is simply $\log_2 K$.
Intuitively, this is the number of bits needed to specify one of $K$ equiprobable
codewords. The second one follows from 
the definition of mutual information 
$I(\b{X};\b{Y})
\equiv H(\b{Y}) - H(\b{Y}|\b{X})$.
For the purpose of finding a lower bound, we consider minimizing the mutual
information per letter instead of the rate, while keeping the von Neumann entropy fixed.
This leads to the following information-theoretical optimization
problem. Given $n$ and the random vector $\b{X}$ as defined above, we wish
to find

     \beq
        R_n(S) =  \frac{1}{n} \,\, \inf_{Q(\b{y}|\b{x}):  S(Q) = S} I(Q) 
        \label{rninf}
     \eeq
 
 where $I(Q)$ is the mutual information 
 
   \beq
      I(Q) =   {\int \!\!\! \int} d\b{x} d\b{y} p(\b{x}) Q(\b{y}|\b{x}) 
      \log \frac{Q(\b{y}|\b{x})}{q(\b{y})}
      =  \int \!\!\! \int d\b{x} d\b{y} q(\b{y}) P(\b{x}|\b{y}) 
      \log \frac{P(\b{x}|\b{y})}{p(\b{x})}
   \eeq

 and 

   \beq
     S(Q) = \frac{1}{n} \int d\b{y} q(\b{y}) S \( \int d\b{x} P(\b{x}|\b{y}) 
     \ket{\mathbf{x}} \bra{\mathbf{x}} \)
   \eeq
   
is the posterior von Neumann entropy, as in (\ref{eqn11}).
The probability density for the marginal $\b{Y}$ distribution is given
by $q(\b{y}) = \int d\b{x} p(\b{x}) Q(\b{y}|\b{x})$ and the conditional
distribution for $\b{X}$ given $\b{Y}$ is 
$P(\b{x}|\b{y}) = p(\b{x}) Q(\b{y}|\b{x}) / q(\b{y})$.
The minimization should be done for a general length $n$ of $\b{x}$. 
We have found a local extremum of this problem \cite{long}, which we conjecture to be global, 
for which the conditional distribution  factorizes, i.e.  
$Q(\b{y}|\b{x}) = \prod_i Q^\la (y_i|x_i)$ 
where 
  
\beq
    Q^\la (y|x) = P^\la (x|y) = \frac{1}{4 \pi} \frac{\la}{e^\la - 1} e^{\la |\langle x \ket{y}|^2 }
    \label{dist}
\eeq
 
 so that $n = 1$ suffices. Here $\la$ plays the role of a Lagrange multiplier.
Some light may be shed on this result by noticing that there are two competing efffects.
One comes from subadditivity of von Neumann entropy, which says that the von Neumann entropy
of the whole is no greater than the sum of the von Neumann entropies of the parts.
This favors large $n$ in order to decrease the von Neumann entropy per letter.
The other comes from superadditivity of mutual information, valid only when $\b{X}$ is
i.i.d. (as in our case) which states that the mutual information between $\b{X}$ and
$\b{Y}$ is no less than the sum of the mutual informations between the corresponding
components $X_i$ and $Y_i$. This favours $n = 1$. The latter effect apparently wins.
The corresponding $R_1(S)$ is parametrized as follows:

\beq
    R_1(\la)  =  \frac{\la}{e^\la - 1} - 1 + \log \(\frac{\la e^\la}{e^{\la} - 1} \)
    \label{ar}
\eeq 
   
\beq
       S(\la) = h_2 \(\frac{1}{\la} - \frac{1}{e^\la - 1}  \)     
       \label{dee}
\eeq
  
where the $\la \in (0,\infty)$ and  
$h_2(p) = -p \log_2 p - (1 - p) \log_2 (1-p)$ is the binary Shannon entropy function.
$R_1(\la)$ is given in nats, and should be converted into bits 
by dividing by $\log 2$. The curve is readily found to be convex, and is shown in Fig 2.

\vspace{7mm} 
\centerline{ {\scalebox{.4}{\includegraphics{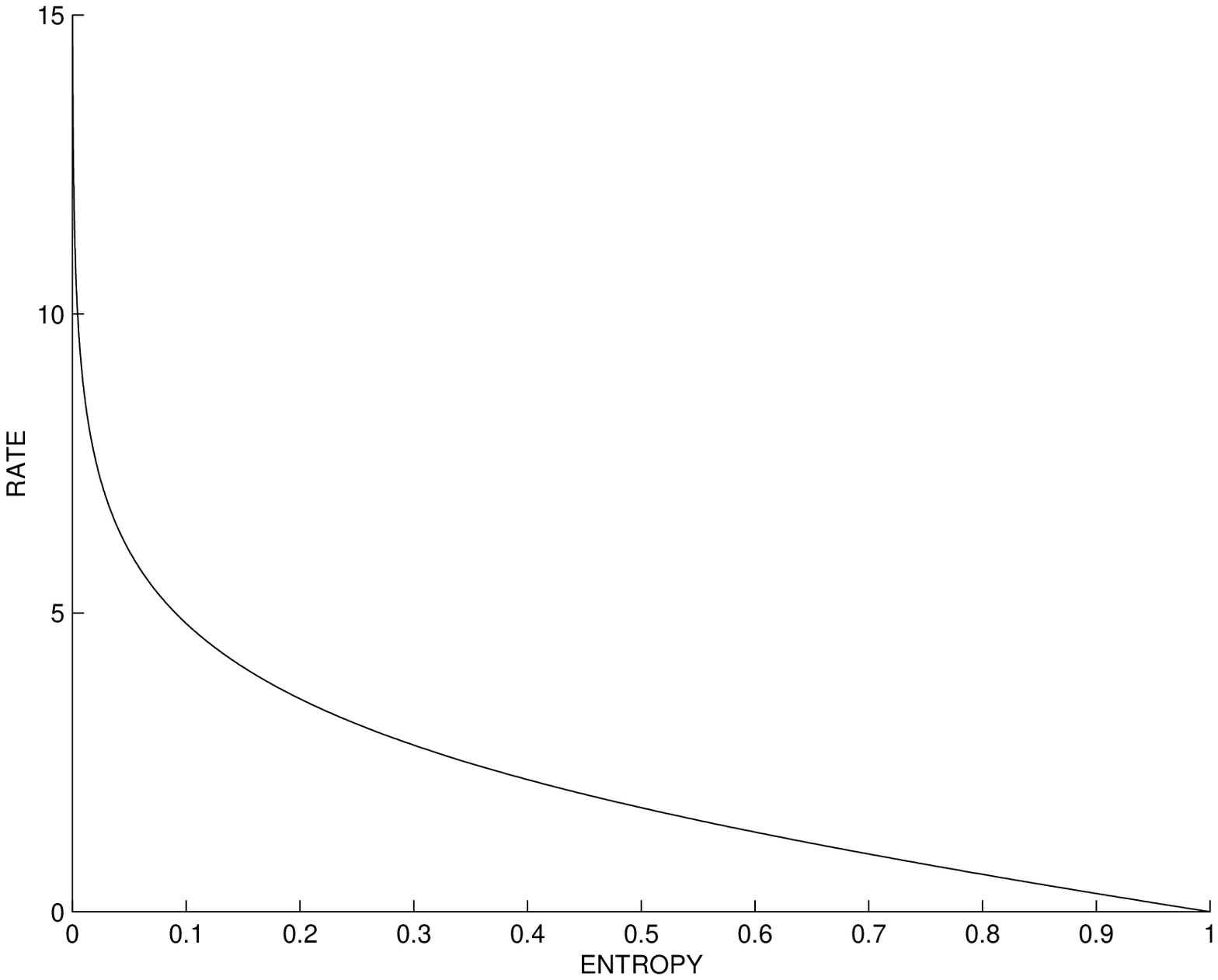}}}}

\centerline{\small {FIG. 2} The rate-entropy function $R(S)$.}

 \vspace{5mm}

So far we have only found a lower bound on $R(S)$. Now we will demonstrate achievability,
and thus establish that $R(S) = R_1(S)$.
It may appear that blocking was not needed after all, but this is due to the fact
that we have not quite solved the coding problem. In particular, our solution
$Q^\la (y|x)$ is not deterministic, as a code should be, but probabilistic. 
Given $x$, $y$ is most likely to be $x$ itself, and then as the arc distance from $x$ increases 
the probability decreases, reaching a minimum at the antipode of $x$. It is only in the 
$\la \rightarrow \infty$ limit that $Q^\la (y|x)$ becomes a $\delta$-function centered at 
$x$, which corresponds to the identity map. This also implies that the second inequality 
in (\ref{eqn2}) is not tight. However, one could expect it to become tight in
the large blocklength limit, since $H(\b{Y})$ is subadditive, and $I(\b{X};\b{Y})$ 
is superadditive. 
The idea is to simulate
the \emph{noisy} single letter channel defined by $P^\la(x|y)$
(acting in the reverse direction, i.e. from $Y$ to $X$) by
the average effect that a {\emph{deterministic}}
coding map (from $\b{X}$ to $\b{Y}$) involving large strings of letters has on the $i$th letter.
To elaborate, let us assume that the $i$th letter in a given codeword $\b{y}$ is 
some $y_i$. Then our code is such that the $i$th components $x_i$ of all
the $\b{x}$'s that get mapped onto $\b{y}$ are distributed as if randomly chosen 
according to the conditional distribution $P^{\la}(x_i|y_i)$. 
Since $P^{\la}(x|y)$ depends only on the overlap $\langle x \ket{y}$, when
Alice rotates $\b{x}$ by the map that sends $\b{y}$ to $\b{0}$,
the block density matrix Bob sees after being told the codeword is
the Schumacher compression friendly tensor product of single qubit density matrices 
$\rho' = \int d{x} P^\la(x|0) \ket{{x}} \bra{{x}}$ with entropy per qubit given by 
$S(\la)$ (\ref{dee}).
The way to construct such a coding map is by using joint typicality decoding,
a technique well known in classical rate-distortion theory \cite{cover}.  
It is necessary first to coarse grain $\CX$ into a disjoint union of small near-circular
caps of diameter $ \approx \ep$  and replace the probability densities
$P^\la(x|y)$ etc. by discrete probabilities $\hat{P}^\la(\hat{x}|\hat{y})$ etc. where  $\hat{x}$ and $\hat{y}$ belong to $\hat{\CX}$, 
the set of cap centroids. 
A \emph{$\de$-typical sequence} $\hat\b{x} \in \hat{\CX}^n$ 
with respect to the distribution ${\hat{p}(\hat{x})}$ is defined as one that satisfies
  
   \beq
    \left | \frac{ N(\hat{a}|\hat{\b{x}})}{n} - \hat{p}({\hat{a}}) \right |
    < \frac{\de}{|\hat{\CX}|} 
    \eeq
   
where $N({\hat{a}}|\hat{\b{x}})$ is the number of occurences of 
$\hat{a} \in \hat{\CX}$ in the sequence $\hat\b{x}$. We call the
set of all such typical sequences the \emph{typical set} $T_\de(\hat{p})$.
In words, a sequence is typical if the fraction of appearances of any given
letter in the sequence approximates the probability for that letter.
Another way of putting it is that picking an element of the sequence at random
approximatley simulates the probability distribution. 
Note that, by the law of large numbers, a sufficiently long sequence chosen according to 
the probability distribution will "almost always" be typical. 
One similarly defines the \emph{jointly typical set}
$T_\de(\hat{P} \hat{q})$ of pairs
of typical sequences $(\hat\b{x}, \hat\b{y}) \in (\hat{\CX} \times \hat{\CX})^n$
with respect to the distribution $\hat{P}^\la(\hat{x}|\hat{y}) \hat{q}(\hat{y})$
\cite{cover}. The coding map is as follows:

\vspace{1mm}

$\bullet$ The codewords $\hat{\b{y}}$ are chosen at random. More precisely, each letter of
each codeword is chosen according to $\hat{q}(\hat{y})$ (which mimics the uniform
distribution). This ensures with high probability that the codewords will be typical of the
distribution $\hat{q}(\hat{y})$.
 
$\bullet$ Mapping a given $\b{x}$ onto a $\hat{\b{y}}$ with the property that the pair
$(\hat{\b{x}},\hat{\b{y}})$ is typical of the joint distribution $\hat{P}^\la(\hat{x}|\hat{y}) \hat{q}(\hat{y})$.
Here $\hat{\b{x}}$ is the componentwise centroid of the cap that contains $\b{x}$.
This implies that if we randomly pick a $\hat{\b{x}}$ and its corresponding $\hat{\b{y}}$,
 the $i$th component pair will equal $(\hat{x}_i,\hat{y}_i)$ with probability 
$\hat{P}^\la(\hat{x}_i|\hat{y}_i) \hat{q}(\hat{y}_i)$. Hence, given $\hat{y}_i$, $\hat{x}_i$ was the source 
letter with probability $P^{\la}(\hat{x}_i|\hat{y}_i)$. This is how the noisy channel
$P^\la(x|y)$ is simulated.
 
The above map fails when there are not enough reproducing codewords
to ensure that one can find a member of the code $B_n$ jointly typical with a given $\hat{\b{x}}$.
It turns out \cite{cover} that the minimal rate for which such an error "almost never" occurs
is precisely the mutual information corresponding to $\hat{P}(\hat{x}|\hat{y}) \hat{q}(\hat{y})$,
which is approximated by $R_1(\la)$ (\ref{ar}). 
Finally, it is necessary to take the $\ep, \de \rightarrow 0$ and $n \rightarrow \infty$
limits carefully to ensure that the pair $(R,S)$ indeed approaches the $R_1(S)$ curve arbitrarily closely
\cite{long}.
Note that joint typicality decoding is suboptimal, strictly speaking. The actual optimal map makes no reference to
coarse graining. The code $B_n$ is chosen at random, and the coding map is the one that minimizes 
$S(B_n)$. However, stating it that way gives us little hope of computing $R(S)$.

Our RSP protocol is now analogous to the simple one described earlier.
Alice wishes to remotely prepare a string of $n$ qubits
using an $(R,S)$ source code. 
She identifies the corresponding codeword and rotates the
original string by the map that sends the codeword  to $\b{0}$ (this is analogous
to mapping the south pole onto the north pole in Lo's example), and 
prepares these qubits in her laboratory.
She may Schumacher compress them without additional blocking to $S n$ qubits.
She teleports these to Bob using $2 S n$ classical bits and 
$S n$ ebits. A further $R n$ bits are sent in order to convey the codeword. Bob reverses Alice's
steps in his laboratory, thus recovering an asymptotically faithful copy
of the qubits to be prepared. The corresponding point in the (b,e)-plane is $(R + 2 S, S)$
per remotely prepared state.
The ebits vs. bits tradeoff curve is shown by the dotted curve in Fig 1. and is parametrized by 
$(R_1(\la) + 2 S(\la), S(\la))$. 

It should be noted that our protocol does not require back-communication,
since it is based on teleportation, which enjoys the same property.  
We conjecture that teleportation based protocols are optimal among all low-entanglement protocols,
and hence that our result is exact.
To show this formally it is crucial to understand the high-entanglement region,
since we expect other candidates to be "generated" by special points in the high-entanglement region
in the same way our upper bound was generated by the teleportation point via $R(S)$.

We are grateful to N.D.Mermin for bringing reference \cite{bennett} to our attention. 
We also thank C.H.Bennett, D.P.DiVincenzo, P.W.Shor, B.M.Terhal and H.-K. Lo for useful discussions 
that revealed misstatements in an earlier version of the paper, and A.K.Pati for pointing us to
reference \cite{pati}. This research was 
supported in part by
the DoD Multidisciplinary University Research
Initiative (MURI) program administered by the Army Research Office under
Grant DAAD19-99-1-0215 and NSF Grant CCR-9980616.

\end{document}